\documentclass[useAMS,usenatbib]{emulateapj}

\usepackage{comment}
\usepackage{xspace}
\usepackage{graphicx}
\usepackage{epsfig}
\usepackage[backref,colorlinks,citecolor=blue]{hyperref}
\usepackage{appendix}

\providecommand{\url}[1]{\href{#1}{#1}}
\providecommand{\dodoi}[1]{doi:~\href{http://doi.org/#1}{\nolinkurl{#1}}}
\providecommand{\doeprint}[1]{\href{http://ascl.net/#1}{\nolinkurl{http://ascl.net/#1}}}
\providecommand{\doarXiv}[1]{\href{https://arxiv.org/abs/#1}{\nolinkurl{https://arxiv.org/abs/#1}}}

\newcommand{\nicer}{\textit{NICER}\xspace}

\shorttitle{Winds in GRS 1915+105 with \nicer}
\shortauthors{Neilsen et al.}

\begin{document}

\title{A Persistent Disk Wind in GRS 1915+105 with \textit{NICER}}

\author{J.\ Neilsen\altaffilmark{1}}
\author{E.\ Cackett\altaffilmark{2}}
\author{R.\ A.\ Remillard\altaffilmark{3}}
\author{J.\ Homan\altaffilmark{4,5}}
\author{J.\ F.\ Steiner\altaffilmark{3}}
\author{K.\ Gendreau\altaffilmark{6}}
\author{Z.\ Arzoumanian\altaffilmark{6}}
\author{G.\ Prigozhin\altaffilmark{3}}
\author{B.\ LaMarr\altaffilmark{3}}
\author{J.\ Doty\altaffilmark{3}}
\author{S.\ Eikenberry\altaffilmark{7}}
\author{F.\ Tombesi\altaffilmark{6,8,9,10}}
\author{R.\ Ludlam\altaffilmark{11}}
\author{E.\ Kara\altaffilmark{8,12,13}}
\author{D.\ Altamirano\altaffilmark{14}}
\author{A.\ C.\ Fabian\altaffilmark{15}}

\altaffiltext{1}{Villanova University, Department of Physics, Villanova, PA 19085, USA; jneilsen@villanova.edu}
\altaffiltext{2}{Wayne State University, Department of Physics \& Astronomy, Detroit, MI 48201, USA}
\altaffiltext{3}{MIT Kavli Institute for Astrophysics and Space Research, Cambridge, MA 02139, USA}
\altaffiltext{4}{Eureka Scientific, Inc., 2452 Delmer Street, Oakland, CA 94602, USA}
\altaffiltext{5}{SRON, Netherlands Institute for Space Research, Sorbonnelaan 2, 3584 CA Utrecht, The Netherlands}
\altaffiltext{6}{NASA Goddard Space Flight Center, Greenbelt, MD 20771, USA}
\altaffiltext{7}{Department of Astronomy, University of Florida, Gainesville, FL 32611, USA}
\altaffiltext{8}{Department of Astronomy, University of Maryland, College Park, MD 20742, USA}
\altaffiltext{9}{Department of Physics, University of Rome ``Tor Vergata,'' Via della Ricerca Scientifica 1, I-00133 Rome, Italy}
\altaffiltext{10}{INAF Astronomical Observatory of Rome, Via Frascati 33, 00078 Monteporzio Catone, Italy}
\altaffiltext{11}{Department of Astronomy, University of Michigan, Ann Arbor, MI 48109, USA}
\altaffiltext{12}{Joint Space Science Institute, University of Maryland, College Park, MD, 20742}
\altaffiltext{13}{NASA Hubble Fellow}
\altaffiltext{14}{School of Physics and Astronomy, University of Southampton, Southampton, SO17 1BJ, UK}
\altaffiltext{15}{Institute of Astronomy, Madingley Road, Cambridge CB3 0HA, UK}

\begin{abstract}
The bright, erratic black hole X-ray binary GRS 1915+105 has long been a target for studies of disk instabilities, radio/infrared jets, and accretion disk winds, with implications that often apply to sources that do not exhibit its exotic X-ray variability. With the launch of \nicer, we have a new opportunity to study the disk wind in GRS 1915+105 and its variability on short and long timescales. Here we present our analysis of 39 \nicer observations of GRS 1915+105 collected during five months of the mission data validation and verification phase, focusing on Fe\,{\sc xxv} and Fe\,{\sc xxvi} absorption. We report the detection of strong Fe\,{\sc xxvi} in 32 ($>80\%$) of these observations, with another four marginal detections; Fe\,{\sc xxv} is less common, but both likely arise in the well-known disk wind. We explore how the properties of this wind depends on broad characteristics of the X-ray lightcurve: mean count rate, hardness ratio, and fractional RMS variability. The trends with count rate and RMS are consistent with an average wind column density that is fairly steady between observations but varies rapidly with the source on timescales of seconds. The line dependence on spectral hardness echoes known behavior of disk winds in outbursts of Galactic black holes; these results clearly indicate that \nicer is a powerful tool for studying black hole winds.
\end{abstract}


\section{Introduction}
\label{sec:intro}

In the 16 years since \citet{L02} first discovered highly-ionized outflowing gas in GRS 1915+105, this remarkable system has played a significant role in expanding our understanding of winds from stellar-mass black holes. Already known for its jets (\citealt{Mirabel94,M98,Fender99}) and bizarre accretion disk variability (e.g., \citealt{B00,K02,Hannikainen05}), GRS 1915+105 embodies much of what we know about the physics of accretion and ejection (see \citealt{FB04}).
\begin{figure*}
\centerline{\includegraphics{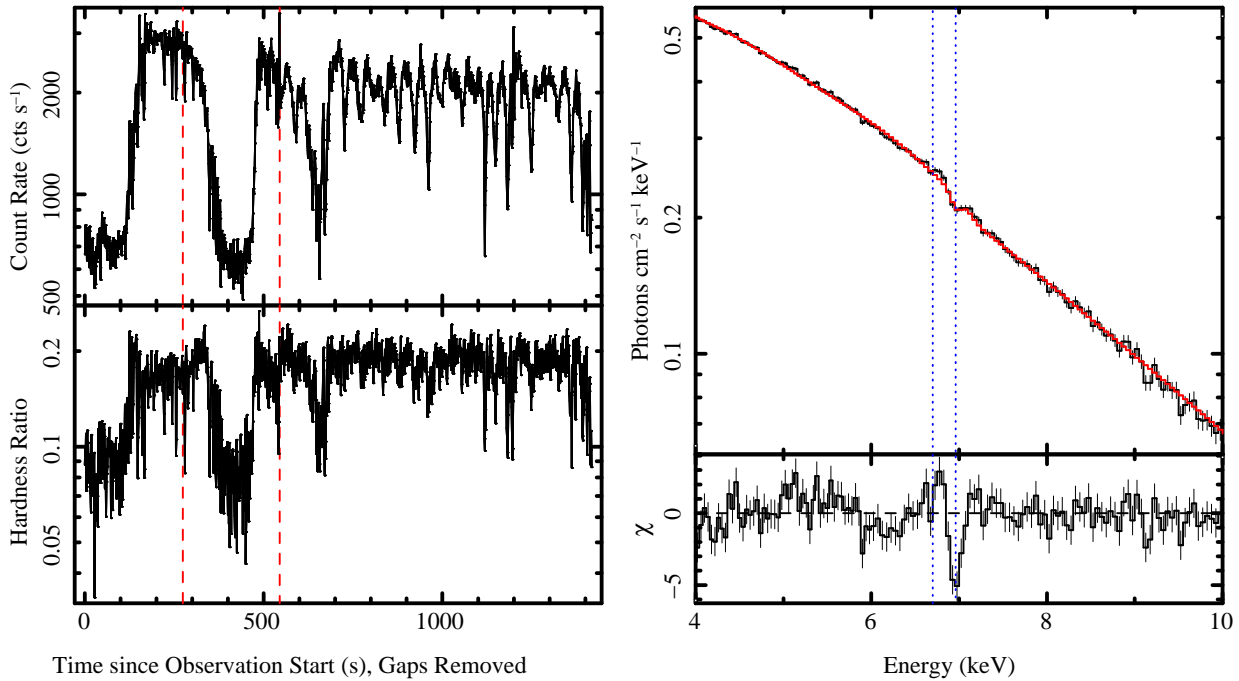}}
\caption{A sample \nicer observation of GRS 1915+105 (ObsID 1103010117). Top left: 1-s lightcurve (0.2--15 keV, not background-subtracted) showing strong erratic variability. Lightcurve gaps (dashed red lines) have been removed for clarity. We also show the hardness ratio (bottom left), here defined as the ratio of the 6--12 keV and 2--4 keV count rates. Right: the corresponding background-subtracted spectrum, rebinned for clarity, with best fit model (red; an absorbed/scattered disk with emission and absorption lines). The locations of the Fe absorption lines are shown with dotted blue lines; residuals are shown with the absorption line normalizations set to zero. See text for details. \label{fig:lcspecwide}}
\end{figure*}

Indeed, the discovery of state-dependent winds in GRS 1915+105 (\citealt{M08,NL09}; see also \citealt{U09,U10,N11a,N12a,Miller16,Zoghbi16}) led to the uncovering of a wind/state dependence for stellar mass black holes in general.  In GRS 1915+105, \citet{NL09} argued that massive winds can reshape the accretion flow to the extent that they suppress or quench jet formation. From archival observations, massive winds are preferentially but not exclusively detected in softer outburst states, where jet emission is generally absent or weak (\citealt{Ponti12,N13a,Homan16} and references therein). Whether this trend reflects the same processes at work in GRS 1915+105 remains an open question, as does the physical origin of winds. Most appear to be consistent with thermal and radiative driving (\citealt{B83,U04,N13a,N16,DT16,Done18}), but there are arguments for MHD processes as well (\citealt{M06b,Fukumura17}). One obstacle to resolving such questions is the scarcity of sensitive spectroscopic monitoring. Without the ability to track the appearance and evolution of black hole winds during outbursts, it is difficult to draw robust conclusions about their life cycles. 

This is one of the many reasons that \nicer (the \textit{Neutron star Interior Composition Explorer}) is so valuable for studies of disk winds: with its spectral resolution (137 eV at 6 keV) and excellent sensitivity, \nicer can detect typical iron absorption lines ($\sim7$ keV) from a highly-ionized wind in just $\sim1$ ks at a flux of $\sim5\times10^{-9}$ erg s$^{-1}$ cm$^{-2}$. Near 1 keV the \nicer X-ray Timing Instrument (XTI) is more sensitive to X-ray absorption lines than the \textit{XMM-Newton} Reflection Grating Spectrometer. And, with its time resolution and flexible scheduling, \nicer can make frequent pileup-free observations of black hole outbursts, enabling the sort of detailed tracking required to understand the physics of winds.
%
%
%
%

Given its contributions to the field, GRS 1915+105 is an ideal first target for \nicer wind studies. Prior spectra of the black hole have previously revealed wind absorption lines in many of its 14 variability classes, but we have little information about the variance in wind properties within and across variability classes (in part due to the limited availability of observations with sufficient spectral resolution).

In this work, we begin the process of remedying this problem with \nicer by investigating the dependence of absorption line behavior on broad lightcurve properties. In Section \ref{sec:obs}, we discuss the dozens of \nicer observations of GRS 1915+105 made in 2017 and our data reduction strategy. In Section \ref{sec:results}, we present our spectral analysis of these observations, focusing on the strengths of the Fe\,{\sc xxv} He$\alpha$ line at 6.7 keV and the Fe\,{\sc xxvi} Ly$\alpha$ line at 7 keV. We discuss the significance of our results in Section \ref{sec:discuss}.


\setcounter{footnote}{0}

\section{Observations and Data Reduction}
\label{sec:obs}

\nicer observed GRS 1915+105 frequently between 2017 June and November, both during commissioning (ObsIDs 0103010101-0103010108) and early Science Operations (ObsIDs 1103010101-1103010134). These 42 ObsIDs had an average (per day) exposure of 1.9 ks. Two ObsIDs had no good exposure time; the others ranged from 123 s to 12 ks. Given \nicer's sensitivity and the high source flux, these observations accumulated $\sim1.4\times10^8$ counts.
%
%

The data were processed using tools from  {\sc nicerdas} (\nicer Data Analysis Software) version 3.0 and the 2018 February 26 release of the \nicer Calibration Database. Specifically, we applied standard calibration and standard time screening with the tools {\tt nicercal} and {\tt nimaketime}, selecting only events that were (1) not flagged as ``overshoot" or ``undershoot" resets (EVENT\_FLAGS=bxxxx00) and (2) detected outside the SAA and at least 30 and 40 degrees above the Earth limb and bright Earth limb, respectively. We created cleaned merged event files with {\tt nicermergeclean} (restricting our energy band to 0.2-15 keV, excluding all non-photon triggers with EVENT\_FLAGS=bx1x000, and using the ``trumpet" filter to eliminate additional known background events). The procedure for rejecting ``hot"\footnote{Here defined as any FPM whose 0--0.2 keV count rate exceeds the FPM ensemble mean by more than $4\sigma$, where $\sigma$ is the sample standard deviation; the mean and $\sigma$ are calculated after excluding the minimum and maximum values from 52 operating FPMs during a given observation.} focal plane modules (FPMs) does not flag any detectors in these particular ObsIDs, but we exclude one observation (1103010118) with a 13--15 keV rate $>4$ cts s$^{-1}$ (see below). The result was 39 observations suitable for analysis. 

We extracted spectra and 1-s lightcurves from these observations using {\sc xselect}. In many cases, the spectra are averaged over significant variability (see Section \ref{sec:discuss}). In order to calculate hardness ratios, we produced lightcurves from the full energy band as well as the 2--4 keV band and the 6--12 keV band (analogous to the $A$ and $B$ bands for \textit{RXTE}).

%
%

\section{Analysis}
\label{sec:results}

A sample lightcurve and spectrum are shown in Figure \ref{fig:lcspecwide}. The lightcurve exhibits the classic erratic variability  of GRS 1915+105, here oscillating between a high flux state and a fainter, softer state. We tentatively identify this as a $\lambda$ state in the classification of \citealt{B00}; a number of other states, including $\rho, \kappa, \gamma, \omega,$ and $\phi$ also appear in our data. The corresponding spectrum is shown from 4--10 keV in the right panel of Figure \ref{fig:lcspecwide}. We note the presence of a small dip in the spectrum at $\sim7$ keV, typical of Fe\,{\sc xxvi} absorption from the known accretion disk wind. In the next subsection, we describe our efforts to model the underlying continuum and the properties of this absorption line. Detailed analysis of the continuum itself is left to other work (J. Steiner, private communication). All of the spectral analysis described below was performed in the Interactive Spectral Interpretation System ({\sc isis}) version 1.6.2-40 (\citealt{HD00}). 

\subsection{Background Subtraction}
As \textit{NICER} is not an imaging instrument, we cannot simply extract a spectrum of the instantaneous X-ray background in the vicinity of the source. Instead, we use background template spectra based on the \textit{RXTE} background fields (\citealt{Jahoda06}; R.\ Remillard, private  communication). The templates are created from cleaned event lists provided by the standard pipeline processing for the \textit{NICER} data archive. The resulting spectra are stacked by 13--15 keV count rate (13 bins from 0--4 cts s$^{-1}$) and generally have tens of ks of exposure.  For a given observation, the 13--15 keV rate determines the background template, which we then rescale using the {\sc isis} functionality {\tt corfile}\footnote{\href{http://space.mit.edu/home/mnowak/isis\_vs\_xspec/back.html}{http://space.mit.edu/home/mnowak/isis\_vs\_xspec/back.html}}. This makes the exact background scaling a free parameter in our spectral fits; background is generally less than a few percent of the 0.25-12 keV count rate.
%
%

\subsection{Continuum Modeling}
For the purpose of detailed photoionization studies of accretion disk winds (\citealt{N11a,N12a,Miller15,K82}) it is necessary to have an accurate physical description of the broadband X-ray spectrum out to $\sim30$ keV. But for simple diagnostics of wind absorption like those we pursue here, we only need to fit the local continuum to sufficient accuracy that we can characterize the line properties. We report our continuum fits but stress that they are designed to fit the data, not for independent physical interpretation.

To this end, we treat the (usually 0.25-12 keV) continuum as absorbed and Comptonized disk emission, described by the model {\tt tbnew*simpl$\otimes$ezdiskbb} (\citealt{Wilms00,Zimmerman05}; {\tt simpl} takes a fraction of its seed photons and scatters them into a power-law-like component; \citealt{Steiner09b}). We use cross-sections from \citet{Verner96} and assume solar abundances from \citeauthor{Wilms00}, with the exception of Si and S, which are free to vary. On top of our continuum model, we include up to four Gaussian components as necessary to account for calibration uncertainties in the 1-3 keV range. Our $\chi^2_{\nu}$ range from 0.94 to 1.4.

\begin{figure*}
\centerline{\includegraphics[width=\textwidth]{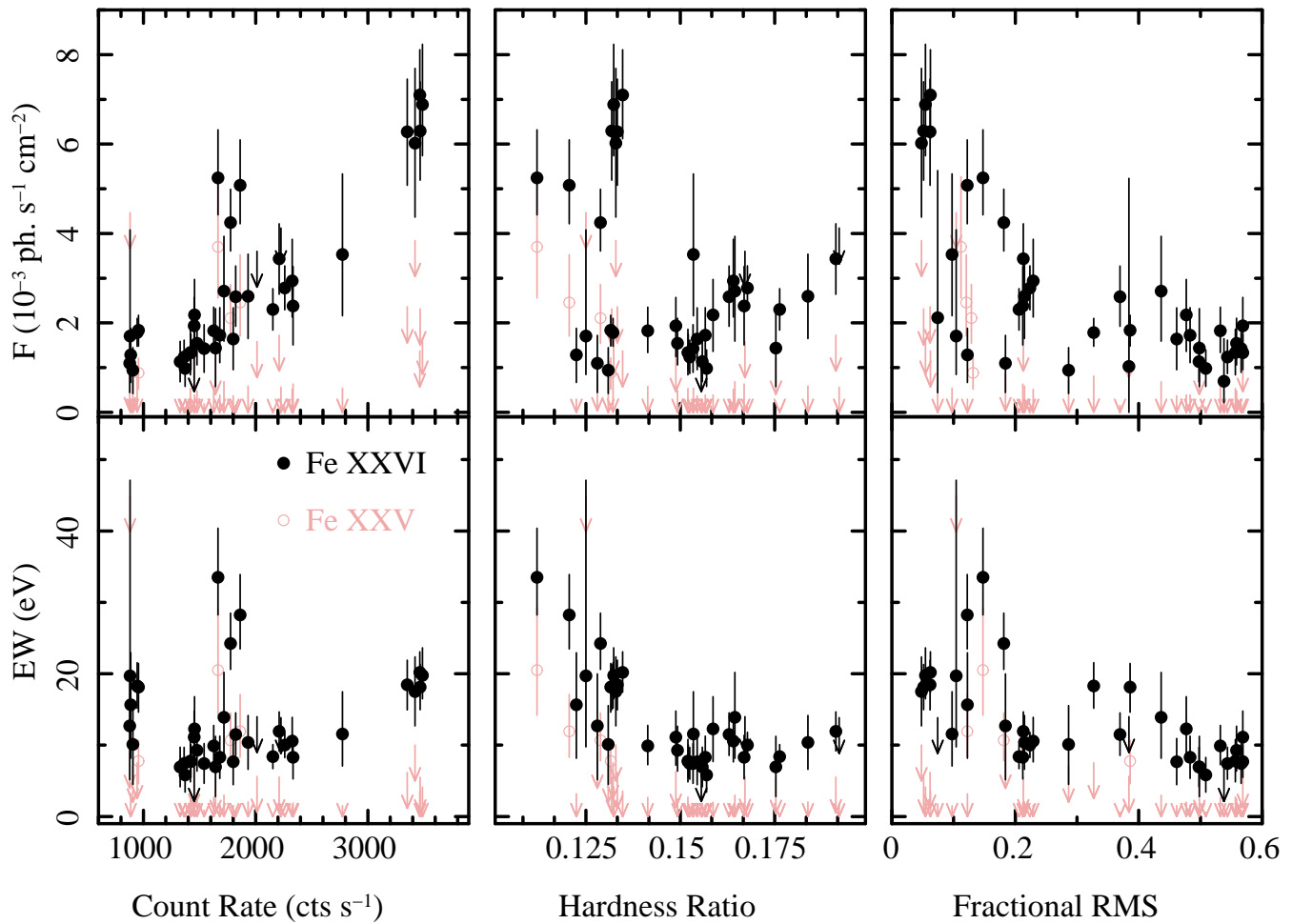}}
\caption{\small Absorption line flux (top) and equivalent width (bottom) vs average count rate (left), 6--12 keV/2--4 keV hardness ratio (middle), and fractional RMS (right) for both Fe\,{\sc xxvi} (black filled circles) and Fe\,{\sc xxv} (pink open circles). Non-detections are shown as 2$\sigma$ upper limits. \label{fig:lineprops}}
%
%
\end{figure*}

The value of the ISM column $N_{\rm H}$ is typically between 6--7.5$\times10^{22}$~cm$^{-2}$, which is reasonable for GRS 1915+105 (e.g., \citealt{L02}). Most observations prefer a Si abundance $A_{\rm Si}\sim1.5-2.5A_{\rm Si, \odot};$ S abundances are generally within 20\% of solar. These may be indicative of excess ISM absorption (e.g., \citealt{L02}) or calibration artifacts. As for the continuum itself, typical disk temperatures are $1.65-2.15$ keV, and none of the observations have a {\tt simpl} scattering fraction that is $>0$ at  90\% confidence. This is not physically significant, since we did not extend the energy range over which {\tt simpl} is computed (which is necessary to properly conserve photons). Furthermore, for highly absorbed sources like GRS 1915+105, fits to soft X-ray spectra (e.g., from the Chandra gratings, with a similar 0.5-10 keV bandpass) can often be fit with models that differ somewhat from the full broadband spectrum (e.g., \citealt{U09}). Assuming a distance of 8.6 kpc (\citealt{Reid14}), the 2--10 keV luminosities range from (3--13)$\times10^{37}$ erg s$^{-1}.$
%
%

\subsection{Line Modeling}
While our continuum model provides a good description of the data overall, many of our observations also show evidence of a broad Fe emission line around 6.4 keV and a narrow absorption line at 7 keV; several observations have numerous line features. To account for the Fe emission line, we include where necessary a Gaussian component between 5.5 and 7 keV with a width $\le0.5$ keV. This is only an approximation to the well-known relativistic lines in GRS 1915+105 (e.g., \citealt{Miller15,Martocchia06}), but it is sufficient given our focus on absorption lines from disk winds.

To model the absorber, we add two\footnote{In a small number of cases, we include additional components to account for Si, S, Ca, and Ni absorption as well.} Gaussian lines to the fit for each observation: one at 6.7 keV (Fe\,{\sc xxv} He$\alpha$, fit between 6.6--6.8 keV) and one at 7 keV (Fe\,{\sc xxvi} Ly$\alpha$, fit between 6.9 and 7.1 keV). For simplicity, we fit these lines with a single Doppler width $\sigma<0.05$ keV. In all but two cases, the lines are unresolved. We fit for the line energy, width, and flux; calculate 90\% confidence intervals; and use the continuum to infer the line equivalent widths (EW; or upper limits). 

%
%
A note on velocities: Though our goal is to study the well-known wind, we sound a brief note of caution regarding the measurement of line energies and blueshifts. The typical statistical uncertainty on our line energies is $\sigma_{\rm E}\sim25$ eV, corresponding to a velocity uncertainty of  $\sim1,100$ km s$^{-1}$ at 90\% confidence. More than half of our observations show blueshifts that are significant at this level, but there is also evidence for systematic shifts in the gain calibration between 5--7 keV that can exceed 20-30 eV. Given these statistical and systematic uncertainties, we cannot robustly infer the presence of an outflow from our data alone. However, \textit{Chandra} grating spectra have measured blueshifted ($\sim400-1700$ km s$^{-1}$) lines in many variability classes of GRS 1915+105 at similar fluxes (e.g., \citealt{L02,M08,NL09,U09,Miller16}), and we proceed under the assumption that the absorption detected here arises in the same wind seen there; an alternative is presented in Section \ref{sec:discuss}.
%
%

\subsection{Line Detections}
Remarkably, we detect signatures of a highly-ionized absorber (Fe {\sc xxvi}) in the vast majority of our observations: 32 detections at $>3\sigma,$ 4 cases of marginal $>2.5\sigma$ detections, and 3 non-detections. These could be due to low continuum signal: the least significant lines are associated with the fewest counts per observation, and with \textit{NICER}'s soft response, most of our counts fall well below 7 keV. Only four observations have  Fe\,{\sc xxv} absorption that is significant at $3\sigma$; the Fe\,{\sc xxvi} line flux is always at least 60\% larger than the Fe\,{\sc xxv} line flux. 

Despite the apparent persistence of the  disk wind, we may ask whether its properties vary with the behavior of the source (e.g., luminosity, spectral shape, or the particular variability class). In our analysis, count rate, hardness ratio (HR=6--12 keV/2--4 keV), and fractional RMS variability (here the RMS/mean of the 1-s lightcurve of an ObsID) act as proxies for these quantities. We focus on Fe\,{\sc xxvi} because we detect it much more frequently than Fe\,{\sc xxv}.

In Figure \ref{fig:lineprops}, we show the absorbed line fluxes and the line EWs vs.\ count rate, hardness ratio, and fractional RMS, for both absorption lines. In the line flux, several trends are clearly apparent.  First, the Fe\,{\sc xxvi} flux increases significantly with count rate (sample correlation coefficient $r=0.83$; \textit{detections} of Fe\,{\sc xxv} follow a similar trend). This is not entirely surprising, as brighter states have more flux to absorb. We account for this using the equivalent width, which measures the line strength relative to the continuum. The lower left panel of Figure \ref{fig:lineprops} shows that the factor of $\sim7$ variability in the line flux is reduced to $\sim2\times$ (with a few outliers) in EW. In other words, the line's EW is a much weaker function of count rate than its flux.

The other trend apparent in the top row of Figure \ref{fig:lineprops} is the decrease in Fe\,{\sc xxvi} line flux with RMS variability. This correlation is more moderate ($r=-0.71$), and may be driven by the detection of strong lines in a soft, steady state (likely similar to the one observed by \citealt{U09}). A moderate anticorrelation ($r=-0.57$) between count rate and RMS may also contribute. We will consider this trend further in Section \ref{sec:discuss}.

The Fe\,{\sc xxvi} line EW also appears to decrease with both hardness ratio and fractional RMS ($r=-0.71$ and $r=-0.61$, respectively). Given the small number of detected Fe\,{\sc xxv} lines, it is difficult to draw conclusions about the line ratio, though this is a useful diagnostic of the ionization parameter (\citealt{L02,U09,N11a,Ponti12}). However, it is worth noting that all of our Fe\,{\sc xxv} detections occur at count rates  $\lesssim2,000$ cts s$^{-1},$ while Fe\,{\sc xxvi} is detected up to nearly 3,500 cts s$^{-1}.$ 
%
%
%
\section{Discussion}
\label{sec:discuss}
In the preceding section, we found a highly-ionized absorber in nearly all \nicer observations of GRS 1915+105, which we attributed to the well-known disk wind. The Fe\,{\sc xxvi}:Fe\,{\sc xxv} line ratios roughly imply ionization parameters $\xi\gtrsim10^4$; this may increase with luminosity. But while the flux of the Fe\,{\sc xxvi} line is a strong function of count rate, its equivalent width is not. 

If these winds are optically thin (see discussion in \citealt{L02}, but also \citealt{N16} for possible counterexamples including GRS 1915+105), this suggests that the Fe\,{\sc xxvi} column density is fairly steady across dozens of observations spanning five months. It could be that the total column density of the wind was steady during this period. This would require the ionization balance to be roughly constant as well, which could happen if (for the observed luminosity range), Fe\,{\sc xxvi} is near its peak ionic abundance, where changes in luminosity produce small changes in the detectable iron column density.  Still, a constant-column wind would be surprising, as prior studies (e.g., \citealt{U09,U10,N11a,N12a}) have found significant variation in wind behavior across states.

In an alternative scenario, there could be large changes in the equivalent hydrogen column that are negated by shifts in the ionization balance (from luminosity, density, or location variations), producing a Fe\,{\sc xxvi} column density that is relatively steady between observations. Saturation could make this scenario less contrived  (\citealt{U09,Miller16,N16}) since the line EW would change slowly with column density. This is not likely to explain our results completely (see, e.g., \citealt{N12a}). Given the radiation field measured by \nicer, photoionization modeling with {\sc xstar} (\citealt{K82}) could in principle probe the plausible parameter space for this wind, but this is beyond the scope of this work.
%
%

Regardless of the details, the pervasiveness of the wind is unexpected. \citet{NL09} argued that the wind and jet in GRS 1915+105 play out a zero-sum game, in which matter is expelled in the wind at the expense of the accretion flow and jet. Did we simply observe an extended interval of wind dominance? Constraints on the jet from coordinated \textit{NICER}/IR observations using CIRCE on the Gran Telescopio Canarias (Dallilar et al., in prep) will soon provide a consistency check on this possibility.

The origins of other trends in Figure \ref{fig:lineprops} are also of interest, especially the decrease in line fluxes and EWs with RMS variability and hardness ratio, respectively. \citet{N11a,N12a} demonstrated that the wind can respond to extreme variability by turning ``on" and ``off" rapidly. To see how this leads to the RMS trend in Figure \ref{fig:lineprops}, consider the observation in Figure \ref{fig:lcspecwide} and suppose that the wind is on during high flux intervals and off during the dips in the lightcurve. The presence of even short periods with no wind will tend to dilute the wind in the average spectrum (as found in one case study by \citealt{N12a}), such that more variable observations might appear to have weaker winds. Time-resolved spectroscopy can reveal the magnitude of this effect, but rapid wind evolution is a plausible explanation for the link between the line flux and RMS variability. We leave a detailed exploration for future work. 

The decrease in the line EW with hardness ratio is notable for mirroring the visibility of winds in black hole outbursts, which preferentially (if not exclusively; \citealt{Homan16}) show winds in softer states. 
The question is whether the decrease in line strength is explained by (a) excess ionization as the spectral hardness increases, or (b) physical variations in the wind between states with different spectra. As yet, there is no convincing answer, either here or from global studies of black hole outbursts, but the fact that Fe\,{\sc xxv} and Fe\,{\sc xxvi} exhibit the most similar and monotonic trends with respect to spectral hardness suggests that spectral state is still one of the best predictors of the presence of winds. 

In the context of state-dependent winds, another question is whether there is any state-to-state variation in the blueshift of the absorber. We do not find any trends in line energy, but nearly half of our detections are statistically consistent with no Doppler shift. Could we be observing a transition from an extended disk atmosphere (e.g, \citealt{Xiang09}) to an outflowing wind? Spectra stacked by state may be able to address this question in the near future. More broadly, with improved calibration and more states observed, a comprehensive spectral analysis of \nicer data (with their high sampling rate and excellent sensitivity) could provide much needed insights for the future. The same strategies applied to \nicer observations of GRS 1915+105 will also be enormously beneficial to monitoring the life cycle of winds in black hole outbursts (see \citealt{N13a}).

\acknowledgements
We thank Jon Miller and the referee for helpful comments. F.T. acknowledges support by the Programma per Giovani Ricercatori - anno 2014 ``Rita Levi Montalcini.'' ACF acknowledges ERC Advanced Grant 340442. This research has made use of data, software and/or web tools obtained from the High Energy Astrophysics Science Archive Research Center (HEASARC), a service of the Astrophysics Science Division at NASA/GSFC and of the Smithsonian Astrophysical Observatory's High Energy Astrophysics Division.

\end{document}